\documentclass[11pt]{article} 
\usepackage[latin1]{inputenc}
\usepackage[T1]{fontenc} 
\usepackage{fullpage}
\usepackage{fancyhdr,lastpage}
\newtheorem{proposition}{Proposition}[section]
\newtheorem{examp}[proposition]{Example}
\newtheorem{lemma}[proposition]{Lemma}
\newtheorem{fact}[proposition]{Fact}
\newtheorem{theorem}[proposition]{Theorem}
\newtheorem{corollary}[proposition]{Corollary}

\newenvironment{proof}{\par\noindent {\it Proof.} \rm}{\ ~~~$\fbox{}$}
\newenvironment{proof2}[1]{\par\noindent {\it Proof of #1.} \rm}{ \ ~~~$\fbox{}$ } 

\begin{document}

\thispagestyle{empty}
\begin{center}
LaRIA~: Laboratoire de Recherche en Informatique d'Amiens\\
Université de Picardie Jules Verne -- CNRS FRE 2733\\
33, rue Saint Leu, 80039 Amiens cedex 01, France\\
Tel : (+33)[0]3 22 82 88 77\\
Fax : (+33)[0]03 22 82 54 12\\
\underline{http://www.laria.u-picardie.fr}
\end{center}

\vspace{7cm}

\begin{center}
\parbox[t][5.9cm][t]{10cm}
{\center

{\bf Quasiperiodic Sturmian words and morphisms

\medskip

F. Levé$^{\rm b}$, G. Richomme$^{\rm a}$
}
\bigskip

\textbf{L}aRIA \textbf{R}ESEARCH \textbf{R}EPORT~: LRR 2006-01\\
(January 2006)
}
\end{center}

\vfill

\hrule depth 1pt \relax

\medskip

\noindent
$^{a b}$ LaRIA, Université de Picardie Jules Verne, \{florence.leve, gwenael.richomme\}@u-picardie.fr

\vspace{-2cm}

\pagebreak

\setcounter{page}{1}
\fancyhf{} 
\fancyfoot[C]{\thepage/\pageref{LastPage}}

\title{Quasiperiodic Sturmian words and morphisms}
\author{F. Levé, G. Richomme\\
LaRIA, Universit\'e de Picardie Jules Verne\\
33, Rue Saint Leu,\\
F-80039 Amiens cedex 1\\
({\tt \{florence.leve,gwenael.richomme\}@u-picardie.fr})
}
\date{\today}
\maketitle

\begin{abstract}
We characterize all quasiperiodic Sturmian words: a Sturmian word is not quasiperiodic if and only if it is a Lyndon word. Moreover, we study links
between Sturmian morphisms and quasiperiodicity.

\medskip
Keywords: Sturmian words, quasiperiodicity, Lyndon words, morphisms
\end{abstract}

\section{Introduction}

The notion of repetition in Strings is central in a lot of researches, in particular in Combinatorics on Words and in Text Algorithms (see for
instance \cite{Lot2002}, \cite{LotIII} for recent surveys).  In this
vein, Apostolico and Ehrenfeucht introduced the notion of
quasiperiodic finite words \cite{AE1993} in the following way: ``a
string $w$ is quasiperiodic if there is a second string $u \neq w$
such that every position of $w$ falls within some occurrence of $u$ in
$w$''.  The reader can consult \cite{AC2001} for a short survey of
studies concerning quasiperiodicity.  In \cite{Mar2004}, Marcus
extends this notion to right infinite words and he opens six
questions. Four of them are answered in \cite{LR2004}.

One of these six questions is: does there exist a non-quasiperiodic Sturmian
word? In \cite{LR2004}, we provide an example of such a word, but this
positive answer is not completely satisfying. Since a first feeling
can be that there exists no (or at most very few) such word, one can
ask for a complete characterization of such non-quasiperiodic Sturmian
words. After some preliminaries in Sections 2, 3 and 4, we provide two answers described below.

Sturmian words have been widely studied because of their many
beautiful properties and links with many fields (see \cite[Chapter 2]{Lot2002} for a recent survey).  One aspect of these words is that
they can be infinitely decomposed over the four morphisms $L_a$, $L_b$, $R_a$ and $R_b$ (see
Section~\ref{secSturmiens} for more details). 
The first characterization of non-quasiperiodic Sturmian words proposed in this paper is based on such a decomposition. More precisely, Theorem~\ref{caracQuasiSturm} states that a Sturmian word is not quasiperiodic if and only if it can be decomposed infinitely over $\{L_a,R_b\}$ or infinitely over $\{L_b,R_a\}$. 

Our second characterization (Theorem~\ref{Lynd0}) provides a more semantic answer: a Sturmian word is not quasiperiodic if and only if it is an infinite Lyndon word.

The proof of our first result uses the fact that some morphisms obtained by compositions of the morphisms $L_a$, $L_b$, $R_a$ and $R_b$ map any infinite words into a quasiperiodic one. We call such a morphism strongly quasiperiodic. 
In Section 7, we characterize the Sturmian morphisms which are strongly quasiperiodic. Let us quote that any Sturmian morphism $f$ is quasiperiodic, that is there exists a non-quasiperiodic word $w$ whose image by $f$ is quasiperiodic.

\section{Generalities}

We assume the reader is familiar with combinatorics on words and
  morphisms (see, e.g., \cite{Lot1983,Lot2002}).  We precise our
  notations.

Given a set $X$ of words (for instance an alphabet A, that is a
non-empty finite set of letters), $X^*$ (resp. $X^\omega$) is the set
of all finite (resp. infinite) words that can be obtained by 
concatenating words of $X$. The empty word $\varepsilon$ belongs to
$X^*$.  The length of a word $w$ is denoted by $|w|$.  By $|w|_a$ we
denote the number of occurrences of the letter $a$ in $w$.  A
finite word $u$ is a \textit{factor} of a finite or infinite word $w$
if there exist words $p$ and $s$ such that $w = pus$. If $p =
\varepsilon$ (resp. $s = \varepsilon$), $u$ is a \textit{prefix}
(resp. \textit{suffix}) of $w$. A word $u$ is a \textit{border} of a
word $w$ if $u$ is both a prefix and a suffix of $w$. A factor $u$ of
a word $w$ is said \textit{proper} if $w \neq u$.

\medskip

Given an alphabet $A$, a(n endo)\textit{morphism} $f$ on $A$ is an
application from $A^*$ to $A^*$ such that $f(uv) = f(u) f(v)$ for any
words $u$, $v$ over $A$.  A morphism on $A$ is entirely defined by the
images of letters of $A$. All morphisms considered in this paper will
be non-erasing: the image of any non-empty word is never empty. The
image of an infinite word is thus infinite and naturally obtained as
the infinite concatenation of the images of the letters of the word.
In what follows, we will denote the composition of morphisms by
juxtaposition as for concatenation of words. Given a set $X$ of
morphisms, we will also note $X^*$ the set of all finite compositions
of morphisms of $X$ and $X^\omega$ the set of all infinite
decompositions of morphisms of $X$.  When a word $w$ is equal to
$\displaystyle \lim_{n \rightarrow \infty} f_1 f_2 \ldots f_n(a)$,
$f_i \in X$, we will say that $w$ can be decomposed (infinitely) over
$X$.

Given a morphism $f$, \textit{powers} of $f$ are defined inductively
by $f^0 = Id$ (the Identity morphism), $f^i = ff^{i-1}$ for integers
$i \geq 1$.  When for a letter $a$, $f(a) = ax$ with $x \neq
\varepsilon$, the morphism $f$ is said \textit{prolongable} on $a$. In
this case, for all $n \geq 0$, $f^n(a)$ is a prefix of
$f^{n+1}(a)$. If moreover, for all $n \geq 0$, $|f^n(a)| <
|f^{n+1}(a)|$, the limit $\displaystyle \lim_{n \rightarrow \infty}
f^n(a)$ is the infinite word denoted $f^\omega(a)$ having all the
$f^n(a)$ as prefixes. This limit is also a fixed point of $f$.

\section{\label{secSturmiens}Sturmian words and morphisms}

 Sturmian words may be defined in many equivalent ways (see
 \cite[chapter 2]{Lot2002} for instance). They are infinite binary
 words.  Here we first consider them as the infinite balanced non
 ultimately periodic words. We recall that a (finite or infinite) word
 $w$ over $\{a,b\}$ is \textit{balanced} if for any factors $u$ and
 $v$ of same length $||u|_a-|v|_a| \leq 1$, and that an infinite word
 $w$ is \textit{ultimately periodic} if $w = uv^\omega$ for some
 finite words $u$ and $v$.

 \medskip

 Many studies of Sturmian words use Sturmian morphisms, that is
 morphisms that map any Sturmian word into a Sturmian word. Séébold \cite{See1991} proved that the set of
 these morphisms is $\{E, L_a, L_b, R_a, R_b\}^*$ where $E, L_a, L_b,
 R_a, R_b$ are the morphisms defined by 

\medskip
$E : \left\{ \begin{array}{l} a \mapsto b \\ b \mapsto a, \end{array} \right.$
$L_a : \left\{ \begin{array}{l} a \mapsto a \\ b \mapsto ab, \end{array} \right.$
$L_b : \left\{ \begin{array}{l} a \mapsto ba \\ b \mapsto b, \end{array} \right.$
$R_a : \left\{ \begin{array}{l} a \mapsto a \\ b \mapsto ba, \end{array} \right.$
$R_b : \left\{ \begin{array}{l} a \mapsto ab \\ b \mapsto b. \end{array} \right.$

Many relations exist between Sturmian words and Sturmian morphisms. For instance, recently the following result was proved:

\medskip
\begin{theorem}{\rm \cite{BHZ2003}}
\label{theoDecSturm}
Any Sturmian word $w$ over $\{a,b\}$ admits a unique representation of the form 
$$\displaystyle w = \lim_{n \rightarrow \infty}
L_a^{d_1-c_1}R_a^{c_1} L_b^{d_2-c_2}R_b^{c_2} \ldots
L_a^{d_{2n-1}-c_{2n-1}} R_a^{c_{2n-1}}
L_b^{d_{2n}-c_{2n}}R_b^{c_{2n}}(a)$$

where $d_k \geq
c_k \geq 0$ for all integer $k\geq 1$, $d_k \geq 1$ for $k \geq 2$ and
if $c_k = d_k$ then $c_{k-1} = 0$.
\end{theorem}

\medskip
Remark: Let us mention that this representation is not expressed as in \cite{BHZ2003} where it is written $$\displaystyle w = T^{c_1}L_a^{d_1}T^{c_2}L_b^{d_2}T^{c_3}L_a^{d_3}T^{c_4}L_b^{d_4}\ldots$$ where $T$ is the shift map defined, for any infinite word $(a_n)_{n \geq 0}$ with $a_n$ letter for any $n \geq 0$, by $T(a_n)_{n \geq 0}=(a_{n+1})_{n \geq 0}$. One can verify that for integers $c$, $d$ such that $d \geq c \geq 0$ and for any infinite word $w$, $T^cL_a^d(w)=L_a^{d-c}R_a^c(w)$ and $T^cL_b^d(w)=L_b^{d-c}R_b^c(w)$. This explains the links between the two representations. The interested reader will also find relations between this representation and the notion of S-adic systems defined by Ferenczi \cite{Fer1999} as minimal dynamical systems generated by a finite number of substitutions.

\medskip

 A particular well-known family of Sturmian words is the set of standard (or
characteristic) Sturmian words. It corresponds to the case where for each $k
\geq 0$, $c_k = 0$. Hence any standard Sturmian word admits a unique
representation on the form:
$$w = \displaystyle \lim_{n \rightarrow \infty} L_a^{d_1} L_b^{d_2}L_a^{d_3} L_b^{d_4}\ldots L_a^{d_{2n-1}} L_b^{d_{2n}}(a)$$ where $d_1 \geq 0$ and $d_k \geq 1$ for all $k \geq 2$. 

\medskip

To end this section, we recall useful relations between Sturmian morphisms.

\begin{theorem}\label{Lot02}
{\rm \cite{Lot2002} (see also \cite{Ric2003c} for a generalization)}
The monoid $\{L_a, L_b, R_a, R_b, E\}^\ast$ of Sturmian morphisms has the following presentation:

\begin{tabular}{ll}
(1) & $EE=Id$,\\
(2) & $E L_a = L_b E$ and $ER_a=R_bE$,\\
(3) & $L_a L_b^n R_a = R_a R_b^n L_a$, for any $n \geq 0$.
\end{tabular}

\end{theorem}

Note that from $(2)$ and $(3)$, we get: $L_bL_a^nR_b=R_bR_a^nL_b$ for any $n \geq 0$.

\section{Word quasiperiodicity and morphisms } 

In this paper, we consider mainly infinite quasiperiodic words.
However we first
recall the notion of finite quasiperiodic words to
allow us some comparisons.

We consider definitions from \cite{AFI1991}.  A word $u$
\textit{covers} another word $w$ if for every $i \in \{1, \ldots,
|w|\}$, there exists $j \in \{1, \ldots, |z|\}$ such that there is an
occurrence of $u$ starting at position $i-j+1$ in the word $w$.
When $u \neq w$, we say that $u$ is a \textit{quasiperiod} of $w$ and that $w$ is \textit{quasiperiodic}.  A
word is \textit{superprimitive} if it is not quasiperiodic
(Marcus \cite{Mar2004} calls {\it minimal} such words).  One can observe
that any word of length 1 is not quasiperiodic.  The word $$w =
abaababaabaababaaba$$ has $aba$, $abaaba$, $abaababaaba$ as
quasiperiods.  Only $aba$ is superprimitive.  More generally in
\cite{AFI1991}, it is proved that any quasiperiodic finite word has exactly
one superprimitive quasiperiod. This is a consequence of the fact that
any quasiperiod of a finite word $w$ is a proper border of $w$.

\medskip

When defining infinite quasiperiodic words, instead of considering the
starting indices of the occurrences of a quasiperiod, for convenience,
we choose to consider the words preceding the occurrences of a
quasiperiod.  An infinite word $\underline{w}$ is
\textit{quasiperiodic} if there exist a finite word $u$ and words
$(p_n)_{n\geq 0}$ such that $p_0 = \varepsilon$ and, for $n \geq 0$,
$0 < |p_{n+1}|-|p_n| \leq |u|$ and $p_nu$ is a prefix of
$\underline{w}$.  We say that $u$ \textit{covers} $\underline{w}$, or
that $\underline{w}$ is $u$-\textit{quasiperiodic}.  The word $u$ is
also called a \textit{quasiperiod} and we say that the sequence
$(p_nu)_{n\geq 0}$ is \textit{a covering sequence of prefixes of the
word $\underline{w}$}.  The reader will find several examples of
infinite quasiperiodic words in \cite{Mar2002,LR2004}. Let us mention
for instance that the well-known Fibonacci word, the fixed point of
the morphism $\varphi$: $a \mapsto ab$, $b \mapsto a$ is
$aba$-quasiperodic.

It is interesting to note that $\varphi^\omega(a)$ has an infinity of
superprimitive quasiperiods (see \cite{LR2004} for a characterization
of all quasiperiods of $\varphi^\omega(a)$). This shows a great
difference between quasiperiodic finite words and quasiperiodic
infinite words. The reader can also note that for any positive integer
$n$, there exists an infinite word having exactly $n$ quasiperiods (as
for example the word $(ab)^na(ab)^\omega)$), or having exactly $n$
superprimitive quasiperiods \cite{LR2004}.

To end this section, let us observe that any quasiperiod of a (finite of infinite) quasiperiodic word $w$ is a prefix of $w$. Hence $w$ has a unique quasiperiod of smallest length that we call the {\it smallest quasiperiod} of $w$. When $w$ is finite, the smallest quasiperiod of $w$ is necessarily its superprimitive quasiperiod. When $w$ is infinite, its smallest quasiperiod is also superprimitive, but there can exist other superprimitive quasiperiods (see above).

Moreover:

\begin{lemma}
\label{lemmaUU}
If $w$ is an infinite quasiperiodic word with smallest quasiperiod $u$, then 
$uu$ is a factor of $w$.
\end{lemma}

\begin{proof}
If $uu$ is not a factor of $w$ then the prefix $v$ of $u$ of length
$|u|-1$ is a quasiperiod of $w$. This is not possible if $u$ is the
smallest quasiperiod.
\end{proof}

\medskip
Let us observe that Lemma \ref{lemmaUU} is not true for finite words as shown by the $aba$-quasiperiodic word $ababa$.

\medskip
In the following we will also use the immediate following fact:

\begin{fact}\label{f(w)QP}
If $w$ is a (finite or infinite) $u$-quasiperiodic word and $f$ is a non-erasing morphism, then $f(w)$ is $f(u)$-quasiperiodic.
\end{fact}

\section{Sturmian non-quasiperiodic words}

In this section, we prove our main result
(Theorem~\ref{caracQuasiSturm}) which is a characterization of all
non-quasiperiodic Sturmian words. Before this, we prove several useful
results.

\medskip

Let $w$ be a Sturmian word. Denoting by $n$ the
least number of $a$ between two consecutive $b$ in $w$ and by $i$ the
initial number of $a$ in $w$, we can deduce from the balance property of $w$ that $w$ belongs to
$a^i\{ba^n, ba^{n+1}\}^\omega$. When $0 < i\leq n$, $w$ belongs to
$\{a^iba^{n-i}, a^iba^{n+1-i}\}^\omega$ and $w$ is
$a^iba^{n-i+1}$-quasiperiodic (and $a^iba^{n-i+1}$ is the smallest
quasiperiod of $w$). Thus:

\begin{fact}
\label{f1}
If $w$ is a non-quasiperiodic Sturmian word, then there exists an
integer $n$ such that $w$ belongs to $a^{n+1}b \{a^nb,
a^{n+1}b\}^\omega \cup \{ba^n, ba^{n+1}\}^\omega$.
\end{fact}

Of course some Sturmian words in $a^{n+1}b \{a^nb, a^{n+1}b\}^\omega
\cup \{ba^n, ba^{n+1}\}^\omega$ are quasiperiodic: it is the case of
the image of any quasiperiodic Sturmian word starting with $a$ by the
Sturmian morphism $L_a^nR_b: a \mapsto a^{n+1}b, b \mapsto a^nb$.

A consequence of Fact~\ref{f1} is:

\begin{lemma}\label{Form4}
For all Sturmian word $w$ and $x \in \{a,b\}$, $L_xR_x(w)=R_xL_x(w)$
is quasiperiodic.
\end{lemma}

\begin{proof}
Without loss of generality, assume $x=a$. From Theorem~\ref{Lot02}, $L_aR_a=R_aL_a$. Let us recall that
$L_aR_a(a) = a$ and $L_aR_a(b) = aba$.
From Fact~\ref{f(w)QP}, if $w$ is a quasiperiodic word, then $L_aR_a(w)$ is
quasiperiodic. Assume now that $w$ is a Sturmian non-quasiperiodic word.
By Fact~\ref{f1}, $w$ belongs to $a^{n+1}b \{a^nb,
a^{n+1}b\}^\omega \cup \{ba^n, ba^{n+1}\}^\omega$ for an integer $n$.
Hence $L_aR_a(w)$ belongs to one of the sets
$a^{n+1}aba \{a^naba,
a^{n+1}aba\}^\omega$ or $\{abaa^n, abaa^{n+1}\}^\omega$. So
$L_aR_a(w)$ is $a^{n+2}ba$-quasiperiodic or
$aba^{n+2}$-quasiperiodic.
\end{proof}

\medskip

Let us observe that $ba^\omega$ and $L_aR_a(ba^\omega) = aba^\omega$ are not quasiperiodic. This shows that Lemma~\ref{Form4} is not true for arbitrary words (even if they are balanced), unlike the next fact which is a direct consequence of the definition of $L_aL_b$: $a \mapsto aba$, $b \mapsto ab$, and $L_bL_a$: $a \mapsto ba$, $b \mapsto bab$.

\begin{fact}
\label{fact3}
For any infinite word $w$, $L_aL_b(w)$ is $aba$-quasiperiodic and 
$L_bL_a(w)$ is $bab$-quasiperiodic.
\end{fact}

\medskip

Lemma~\ref{Form4} and Fact~\ref{fact3} will be useful to prove that our condition in Theorem~\ref{caracQuasiSturm} is necessary. To show it is sufficient, we now consider situations where the image of a word by a Sturmian
morphism is not necessarily quasiperiodic.

\begin{lemma}
\label{l4bis}
Let $x \in \{a,b\}$ and let $w$ be a balanced word starting with $x$.
The word $L_x(w)$ is quasiperiodic if and only if $w$ is
quasiperiodic. Moreover in this case, the smallest quasiperiod of $L_x(w)$ is the word $L_x(v)$ where $v$ is the smallest quasiperiod of $w$.
\end{lemma}

\begin{proof} 
Without loss of generality, we consider here that $x = a$. 

From Fact~\ref{f(w)QP}, if $w$ is quasiperiodic then $L_a(w)$
is quasiperiodic.

From now on we assume that $L_a(w)$ is $u$-quasiperiodic where $u$ is
the smallest quasiperiod of $L_a(w)$.  If $w$ has at most one
occurence of $b$, then $w = a^\omega$ or $w = a^nba^\omega$ for an
integer $n \geq 0$. Since $L_a(w)$ is quasiperiodic, we have
$w=a^\omega$ and we verify that the smallest quasiperiod of $w$ and
$L_a(w)$ is $a = L_a(a)$.  From now on we assume that $w$ contains at
least two occurrences of the letter $b$. Denoting by $n$ the least
number of $a$ between two consecutive occurrences of $b$ in $w$ and by
$i$ the number of $a$ before the first $b$, since $w$ is balanced, $w
\in a^i\{ba^n, ba^{n+1}\}^\omega$ and $0 \leq i \leq n+1$.

If $0 < i \leq n$, then $w$ and $L_a(w)$ are quasiperiodic with
respective smallest quasiperiod $a^iba^{n-i+1}$ and
$a^{i+1}ba^{n-i+1}=L_a(a^iba^{n-i+1})$.

By hypothesis, $w$ starts with $a$, so we cannot have $i=0$. 

In the case $i=n+1$: $w \in a^{n+1}b\{a^nb, a^{n+1}b\}^\omega$ and
$L_a(w) \in a^{n+2}\{ba^{n+1}, ba^{n+2}\}^\omega$.  Since $u$ is a
quasiperiod of $L_a(w)$, $u$ is a prefix of $L_a(w)$ and starts with
$a^{n+2}b$. By Lemma~\ref{lemmaUU}, $uu$ is a factor of $L_a(w)$. It
follows that $u$ ends with $b$ and $u = L_a(v)$ for a word $v \in
\{a^{n}b,a^{n+1}b\}^\ast$. Now we prove that $v$ is a quasiperiod of
$w$. Let $(p_ku)_{k\geq 0}$ be a covering sequence of $L_a(w)$ ($p_0 =
\varepsilon$ and for all $k \geq 0$, $p_ku$ is a prefix of $L_a(w)$
and $|p_{k+1}|-|p_k| \leq |u|$).  Since $u$ starts with $a^{n+2}b$,
for each $k \geq 0$, there exists a word $p_k'$ such that $p_k =
L_a(p_k')$. Of course, $p_0' = \varepsilon$.  Since $v \in \{a^{n}b,
a^{n+1}b\}^\ast$, we can deduce for each $k \geq 0$ that $p_k'v$ is a
prefix of $w$. If for a $k$, $|p_{k+1}'| - |p_k'| > |v|$, then 
$p_{k+1}' = p_k'vy$ for a word $y$ and consequently $p_{k+1} =
p_kuL_a(y)$ which contradicts the fact that $|p_{k+1}| - |p_k| \leq
|u|$. So for each $k \geq 0$, $|p_{k+1}'| - |p_k'| \leq |v|$. We have
shown that $(p'_kv)_{k \geq 0}$ is a covering sequence of $w$, so $v$
is a quasiperiod of $w$.  Assume $w$ has a quasiperiod $v'$ strictly
smaller than $v$. Both $v$ and $v'$ are prefixes of $w$, so $v=v's$
for a non-empty word $s$. Then $|L_a(v')|=|L_a(v)|-|L_a(s)|<|L_a(v)|$
and $L_a(v')$ is a quasiperiod of $L_a(w)$ strictly smaller than
$u=L_a(v)$. This contradicts the definition of $u$, so $v$ is the
smallest quasiperiod of $w$.

\end{proof}

\medskip

\begin{lemma}
\label{l4ter}
Let $x, y$ be letters such that $\{x, y\} = \{a,b\}$ and let $w$ be a
word starting with $x$.  The word $R_y(w)$ is
quasiperiodic if and only if $w$ is quasiperiodic. Moreover when these words are quasiperiodic, the smallest quasiperiod of $R_y(w)$ is the word $R_y(v)$ where $v$ is the smallest quasiperiod of $w$.
\end{lemma}

\begin{proof}
Without loss of generality, we consider here that $x = a$ and $y = b$. 

From Fact~\ref{f(w)QP}, if $w$ is quasiperiodic then $R_b(w)$
is quasiperiodic.

Assume now that $R_b(w)$ is quasiperiodic and let $u$ be its smallest
quasiperiod.  By hypothesis, $w$ starts with $a$, so does $u$. Since
$aa$ is not a factor of $R_b(w)$ whereas by Lemma~\ref{lemmaUU} $uu$
is a factor of $R_b(w)$, we deduce that $u$ ends with $b$. Thus there
exists a word $v$ such that $u = R_b(v)$.  As done in the proof of
Lemma~\ref{l4bis} for the case $w \in a^{n+1}\{ba^n, ba^{n+1}\}^\omega$, we can show that $v$ is a quasiperiod of $u$ and more precisely that it is its smallest quasiperiod.
\end{proof}

\medskip

The reader can observe one difference between the two previous lemmas:
Lemma~\ref{l4bis} considers only balanced words when
Lemma~\ref{l4ter} works with arbitrary words (starting with $x$).
Note that Lemma~\ref{l4bis} becomes false if we do not consider
balanced words. Indeed the word $w = abab(aaab)^\omega$ is not
quasiperiodic, whereas $L_a(w) = aabaabaa(aabaa)^\omega$ is
$aabaa$-quasiperiodic.  The two lemmas become also false if we
consider Sturmian words starting with $y$ where $\{x,y\}=\{a,b\}$. 
Indeed, let us consider the case $x=a$, $y=b$: it is known
\cite{LR2004} that the word $w = (L_bR_a)^\omega (a)$ is not
quasiperiodic; this Sturmian word starts with $b$ and the word
$L_a(w)$ (resp. $R_b(w)$) is $aba$-quasiperiodic
(resp. $bab$-quasiperiodic).

\medskip

We can now establish the announced characterization of
non-quasiperiodic Sturmian words.

\begin{theorem}
\label{caracQuasiSturm}
A Sturmian word $w$ is not quasiperiodic if and only if it can be
infinitely decomposed over $\{L_a, R_b\}$ or over
$\{L_b, R_a\}$.
In other words a Sturmian word $w$ is not quasiperiodic if and only if
$$\displaystyle w = \lim_{n\rightarrow\infty} L_a^{d_1}R_b^{d_2} L_a^{d_3} R_b^{d_4}\ldots L_a^{d_{2n-1}} R_b^{d_{2n}}(a)$$ or
$$\displaystyle w = \lim_{n\rightarrow\infty} L_b^{d_1}R_a^{d_2} L_b^{d_3} R_a^{d_4}\ldots L_a^{d_{2n-1}} R_b^{d_{2n}}(a)$$ 
where $d_k \geq 1$ for all $k\geq 2$ and $d_1 \geq 0$.
\end{theorem}

\begin{proof}
We first show that the condition is necessary.
Let $w$ be a non-quasiperiodic Sturmian word.
By Theorem~\ref{theoDecSturm}, 
$$\displaystyle w = \lim_{n\rightarrow\infty} L_a^{d_1-c_1}R_a^{c_1} L_b^{d_2-c_2}R_b^{c_2} \ldots L_a^{d_{2n-1}-c_{2n-1}} R_a^{c_{2n-1}} L_b^{d_{2n}-c_{2n}}R_b^{c_{2n}}(a)$$ where $d_k \geq
c_k \geq 0$   for all integer $k\geq 1$, $d_k \geq 1$
for $k \geq 2$ and if $c_k = d_k$ then $c_{k-1} = 0$.

By Lemma~\ref{Form4}, for $x \in \{a, b\}$ and any Sturmian word,
$L_xR_x(w)$ is quasiperiodic. By Fact~\ref{f(w)QP}, this implies that for all $k \geq 1$,
$c_k = d_k$ or $c_k = 0$.

Assume that $c_k = 0$ and $c_{k+1} = 0$ for an integer $k \geq 1$.
Then $w = fL_aL_b(w')$ or $w = fL_bL_a(w')$  for a Sturmian word $w'$ and a morphism $f$.
By Fact~\ref{fact3}, $w$ is quasiperiodic. 
So for each $k \geq 1$, $c_k = 0$ implies $c_{k+1} = d_{k+1}$. 

We know that for each $k \geq  2$, $c_k=d_k$ implies $c_{k-1}=0$. This is equivalent to say that for each $k \geq 1$, $c_k \neq 0$ implies $c_{k+1} \neq d_{k+1}$. 
But there for each $k$, $c_k=d_k$ or $c_k=0$. Thus $c_k=d_k$ implies $c_{k+1}=0$, the condition is necessary.

\medskip

Let us now show that any Sturmian word $w$ that can be decomposed infinitely over
$\{L_a, R_b\}$ is not quasiperiodic (case $\{L_b, R_a\}$ is
similar). Assume by contradiction that it is not the case. Let ${\cal S}$ be
the set of all Sturmian words $w$ that can be decomposed over $\{L_a,
R_b\}$ and that are quasiperiodic. Let $u$ be a quasiperiod of
smallest length among all quasiperiods of words in ${\cal S}$, and let $w$ be
an element of ${\cal S}$ having $u$ as quasiperiod. By definition, $w = L_a(w')$ or $w = R_b(w')$ for a
word $w'$ in ${\cal S}$. Since $d_3 \neq 0$, $w$ starts with the letter $a$. By Lemmas~\ref{l4bis} and ~\ref{l4ter}, $u =
L_a(v)$ or $u = R_b(v)$ where $v$ is the smallest quasiperiod of
$w'$. Since $a^\omega$ and $b^\omega$ are not Sturmian words (they are
balanced but not ultimately quasiperiodic), $|v|_a \neq 0$ and $|v|_b
\neq 0$. Consequently $|v| < |u|$. This contradicts the choice of
$u$. Hence ${\cal S}$ is empty.
\end{proof}

\medskip

Given a word $w$, let us denote $X(w)$ the set of infinite words having the same set of factors than $w$: $X(w)$ is invariant by the shift operator and is called the subshift associated with $w$. When $w$ is Sturmian, it is known (see \cite{BHZ2003}) that a word $w'$ belongs to $X(w)$ if and only if it is Sturmian and the associated sequence $(d_k)_{k \geq 0}$  in its decomposition of Theorem~\ref{theoDecSturm} is the same as the one involved in the decomposition of $w$.

To end this section, we observe that any standard Sturmian word (that is a Sturmian word that can be decomposed
using only $L_a$ and $L_b$) is necessarily quasiperiodic. This gives a
new proof of a result by T.~Monteil \cite{Mon05a, Mon05b}: any Sturmian subshift  contains a
quasiperiodic word (let us mention that the resutl of T. Monteil is more precisely: any Sturmian subshift contains a multiscaled quasiperiodic word, that is a word having an infinity of quasiperiods). The interested reader will find  materials in Section \ref{section_sturm} to show that any standard Sturmian word has an infinity of quasiperiods (see Lemma~\ref{Form1}). Theorem~\ref{caracQuasiSturm} also shows that in any Sturmian subshift, there is a {\it non}-quasiperiodic word.

\section{A connection with Lyndon words}

The aim of this short section is to give another characterization of non-quasiperiodic Sturmian words related to Lyndon words (see Theorem~\ref{Lynd0} below).

Let us recall notions on finite \cite{Lot1983} and infinite \cite{SMDS1994} Lyndon words . We call {\it suffix} of an infinite word $w$ any word $w'$ such that $w=uw'$ for a given word $u$. When $u \neq \varepsilon$, we say that $w'$ is a {\it proper suffix} of $w$. This definition allows us to adopt the same definition for finite and infinite Lyndon word. Let $\preceq$ be a {\it total order} on $A$ (in what follows, $\{a \prec b\}$ denotes the alphabet $\{a,b\}$ with $a \prec b$). This order can be extended into the lexicographic order on words over $A$. A (finite or infinite) word over $(A,\preceq)$ is a {\it Lyndon word} if and only if $w$ is strictly smaller than all its proper suffixes. Any infinite Lyndon word has infinitely many prefixes that are (finite) Lyndon words (and so an infinite Lyndon word can be viewed as a limit of these prefixes). The following basic property of finite Lyndon words was pointed out by J.P. Duval (see Acknowledgments):

\begin{fact}\label{LyndBorders}
Any finite Lyndon word is unbordered, that is the only borders of a Lyndon word $w$ are $\varepsilon$ and $w$.
\end{fact}

This allows us to state a relation between infinite Lyndon words and non-quasiperiodic infinite words (cf Corollary~\ref{Lynd2}).

\begin{fact}\label{Lynd1}
If $w$ is an infinite $u$-quasiperiodic word, then any prefix of $w$
of length at least $|u|+1$ is not unbordered.
\end{fact}

\begin{proof}
If $p$ is a prefix of $w$ of length at least $|u|+1$, then $p$ has for
suffix a prefix $s$ of $u$ (of length at most $|u|$). Since $u$ is a
prefix of $w$, $u$ is also a prefix of $p$, and so $s$ is a border of
$p$.
\end{proof}

\begin{corollary}\label{Lynd2}
Any Lyndon word is not quasiperiodic.
\end{corollary}

Our main Theorem~\ref{Lynd0} is a direct consequence of this corollary and the following characterization. Following~\cite{Ric2003b} we say that a morphism $f$ {\it preserves (finite) Lyndon words} if for any (finite) Lyndon word $u$, $f(u)$ is also a Lyndon word. We have:

\begin{proposition}{\rm \cite{Ric2003b}}\label{CaracSturmianLyndonMorphisms}
A Sturmian morphism $f$ preserves Lyndon words over $\{a \prec b\}$ if and only if $f \in \{L_a,R_b\}^\ast$.
\end{proposition}

\begin{theorem}\label{Lynd0}
A Sturmian word $w$ over $\{a,b\}$ is non-quasiperiodic if and only if $w$ is an infinite Lyndon word over $\{a \prec b\}$ or over $\{b \prec a\}$.
\end{theorem}

\begin{proof}
Let $w$ be a Sturmian word. By corollary~\ref{Lynd2}, if $w$ is an infinite Lyndon word then $w$ is not quasiperiodic.

Assume now that $w$ is not quasiperiodic. By Theorem~\ref{caracQuasiSturm}, $\displaystyle w=\lim_{n\rightarrow\infty}L_a^{d_1}R_b^{d_2}\ldots L_a^{d_{2n-1}}R_b^{d_{2n}}(a)$ or $\displaystyle w=\lim_{n\rightarrow\infty}L_b^{d_1}R_a^{d_2}\ldots L_b^{d_{2n-1}}R_a^{d_{2n}}(a)$ for some integers $(d_k)_{k \geq 1}$ such that $d_k \geq 1$ for all $k \geq 2$ and $d_1 \geq 0$.
Proposition~\ref{CaracSturmianLyndonMorphisms} implies that, since $a$ is a Lyndon word, for each $n \geq 1$, $L_a^{d_1}R_b^{d_2}\ldots L_a^{d_{2n-1}}R_b^{d_{2n}}(a)$ is a Lyndon word over $a \prec b$ and $L_b^{d_1}R_a^{d_2}\ldots L_b^{d_{2n-1}}R_a^{d_{2n}}(a)$ is a Lyndon word over $b \prec a$. Hence $w$ is an infinite Lyndon word over $a \prec b$ or over $b \prec a$. 
\end{proof}

\medskip

To end this section we study the converse of Corollary~\ref{Lynd2} and Fact~\ref{Lynd1}.

The converse of Corollary~\ref{Lynd2} is not true in general. For instance we can consider any Sturmian word $w$ over $\{a,b\}$ and the word $p=ababaaa$. Then $pw$ is not quasiperiodic since $p$ is not balanced and so not a factor of $w$. Moreover, since $p$ starts with the letter $a$, $pw$ cannot be a Lyndon word if $b \prec a$. It is neither a Lyndon word if $a \prec b$ since for any prefix $p'$ of $w$, $aaap' \prec w$.

The converse of Fact~\ref{Lynd1} is also false: let $w$ be an infinite word and $p$ be an integer, if all prefixes of $w$ of length greater than $p+1$ are unbordered, then $w$ is not necessarily quasiperiodic. To prove this, it is sufficient to consider the word $w=aba^\omega$.

A more complex but interesting example, pointed out by P. S\'e\'ebold (see Aknowledgements), is the well-known Thue-Morse
word ${\bf T}$, fixed point of the morphism $\mu$ such that
$\mu(a)=ab$ and $\mu(b)=ba$. The word ${\bf T}$ starts with $abb$ and
any prefix of length at least $4$ ends with $a$, $ab$ or $abb$. But
${\bf T}$ is not quasiperiodic: indeed it is well-known that ${\bf T}$
is overlap-free (a word is overlap-free if it contains no factor of
the form $xuxux$ where $x$ is a letter, or equivalently it contains no
factor that can be written both $pv$ and $vs$ with $|p|<|v|$) and we
can observe that:

\begin{fact}\label{overlap-free}
An overlap-free infinite word is never quasiperiodic.
\end{fact}

\begin{proof}
Let $w$ be a $u$-quasiperiodic infinite word and let $(p_nu)_{n \geq
0}$ be a covering sequence of $w$. If there exists $n \geq 0$ such
that $|p_{n+1}|-|p_n|<|u|$, then $p_{n+1}u=p_nus$ for a word $s$ such
that $s=|p_{n+1}|-|p_n|<|u|$. Hence there exists a word $p$ such that
$us=pu$, then $w$ is not overlap-free. If for all $n \geq 0$ we have
$|p_{n+1}|-|p_n|=|u|$, then $w=u^\omega$ is also not overlap-free.
\end{proof}

\medskip
Finally let us mention that this fact is not valid for finite words since there exist some overlap-free words that are square (see \cite{Thu1912}, cf. also \cite{Ber1992} for a characterization of such words).

\section{Sturmian morphisms and quasiperiodicity}\label{section_sturm}

We say that a morphism $f$ is \textit{quasiperiod-free} if for any
non-quasiperiodic word $w$, $f(w)$ is also non-quasiperiodic. A
non-quasiperiod-free morphism will just be called
\textit{quasiperiodic}.  Let us observe that all Sturmian morphisms
(except $E$ and $Id$) are quasiperiodic. To verify it, it is
sufficient to show that $L_a$, $L_b$, $R_a$ and $R_b$ are
quasiperiodic. For $L_a$ and $R_a$ (case $L_b$ and $R_b$ are similar)
we have: $aba^\omega$ and $ab^\omega$ are non-quasiperiodic although
$L_a(aba^\omega) = aba(ab)^\omega$ and $R_a(ab^\omega) = a(ba)^\omega$
are $aba$-quasiperiodic.

In the previous section, we encounter (Lemma~\ref{Form4} and
Fact~\ref{fact3}) two different kinds of Sturmian morphisms. The
morphism $L_aL_b$ maps any word into a quasiperiodic one, whereas
there exists a non-quasiperiodic word $w$ such that $L_aR_a(w)$ is not
quasiperiodic. Generalizing these two examples we observe that the set
of quasiperiodic morphisms can be partitioned using the following
notions:

\begin{enumerate}
\item
A morphism $f$ on $A$ is called {\em strongly quasiperiodic} (resp. on
a subset $X$ of $A^\omega$) if for each non-quasiperiodic infinite
word $w$ (resp. $w \in X$), $f(w)$ is quasiperiodic.
\item
A morphism $f$ on $A$ is called {\em weakly quasiperiodic} (resp. on
a subset $X$ of $A^\omega$) if there
exist two non-quasiperiodic infinite words $w, w'$ (resp. $w, w' \in X$)
such that $f(w)$ is quasiperiodic, and $f(w')$ is non-quasiperiodic.
\end{enumerate}

The aim of this section is to answer the two following questions:
\begin{itemize}
\item Which are the strongly (resp. weakly) quasiperiodic Sturmian morphisms?
\item Which are the strongly (resp. weakly) quasiperiodic Sturmian morphisms on (the set of) Sturmian words?
\end{itemize}

We note that the two questions have different answers. Indeed $L_aR_a$
as shown by Lemma~\ref{Form4} is strongly quasiperiodic on Sturmian
words, but as already said, $L_aR_a(ba^\omega)$ is not quasiperiodic.
Of course, any strongly quasiperiodic Sturmian morphism is strongly
quasiperiodic on Sturmian words, or equivalently (since a Sturmian
morphism is quasiperiodic), any weakly quasiperiodic Sturmian morphism on
Sturmian words is weakly quasiperiodic.

\subsection{A property of strongly quasiperiodic morphisms}

Before going further, we mention the following immediate result:

\begin{lemma}\label{comp}
Let $f$ be a morphism. If there exist morphisms $f_1$, $f_2$, $f_3$
such that $f=f_1f_2f_3$ and such that $f_2$ is strongly quasiperiodic, then
$f$ is strongly quasiperiodic.
\end{lemma}

We observe that (quite naturally) Lemma~\ref{comp} becomes false
when replacing strongly quasiperiodic by weakly quasiperiodic.  For
instance, taking $f_1 = Id$, $f_2 = L_a$ and $f_3 = L_b$, we have
$f_2$ weakly quasiperiodic and $f_1f_2f_3$ strongly quasiperiodic.
There are cases where we can have $f_2$ weakly quasiperiodic and
$f_1f_2f_3$ quasiperiod-free, but this is not possible when $f_1$, $f_2$
and $f_3$ are Sturmian morphisms since all Sturmian morphisms are
quasiperiodic. To give an example of such a case, we need the following result:

\begin{lemma}
The morphism $g$ defined by $g(a) = abab$ and $g(b) = aaaa$ is a quasiperiod-free morphism.
\end{lemma}

\begin{proof}
Let $w$ be an infinite word such that $g(w)$ is quasiperiodic. We show
that $w$ is also quasiperiodic. Let $u$ be the smallest quasiperiod of
$g(w)$. Since $u$ is a prefix of $g(w)$, $u = g(v)p$ for a proper
prefix $p$ of $g(a) = abab$ or of $g(b) = aaaa$: $p \in \{\varepsilon,
a, aa, aaa, ab, aba\}$.  First we observe that if $a$ or $b$ does not occur in
$w$, then $w$ is quasiperiodic.  From now on we assume that both $a$
and $b$ occur in $w$. Consequently $|v|_a \neq 0$ and $|v|_b \neq 0$.
It follows that $g(v)$ starts with $(ab)^{2n}aaaa$ for an integer $n
\geq 0$ and with $a^{4m}abab$ for an integer $m \geq 0$: of course $m = 0$ or $n
=0$.  Moreover $g(v)$ ends with $aaaa(ab)^{2n'}$ for an integer $n'
\geq 0$ and with $ababa^{4m'}$ for an integer $m' \geq 0$: once again $m' = 0$ or
$n' =0$.  By Lemma~\ref{lemmaUU}, $uu$ is a factor of $g(w)$. We then
deduce that $p =\varepsilon$ since for all the other potential values, none of the words in $\{aaaa(ab)^{2n'},ababa^{4m'}\}p\{(ab)^{2n}aaaa,a^{4m}abab\}$ could be a factor of $g(w)$. Let $(p_lu)_{l\geq 0}$ be a covering
sequence of prefixes of $g(w)$. As done in the proof of
Lemma~\ref{l4bis}, we can find a covering sequence
$(p_l'v)_{l\geq 0}$ of prefixes of $w$: the word $v$ is a quasiperiod
of $w$.
\end{proof}

\medskip
Now let us consider the morphisms $f_1 = Id$, $f_2 = L_a$, and $f_3$
defined by $f_3(a) = bb$, $f_3(b) = aaaa$. By the previous lemma $f_1f_2f_3 = g$ is
quasiperiod-free whereas $f_2$ is weakly quasiperiodic.

To end this section, we let the reader verify that $f_3$ is quasiperiod-free and more generally that any morphism $h$ defined by $h(a)
= a^i$, $h(b) = b^j$ with $i \geq 1$ and $j \geq 1$ is quasiperiod-free.

\subsection{\label{secMorphismesQuasiSturms}Weakly and strongly
quasiperiodic Sturmian morphisms}

In this section, we characterize weakly quasiperiodic Sturmian
morphisms. (Equivalently this characterizes strongly quasiperiodic Sturmian morphisms since any Sturmian morphism is weakly or strongly quasiperiodic.)

\begin{proposition}
\label{caracStronglyQuasiperiodicMorphism}
A Sturmian morphism is weakly
quasiperiodic if and only if it belongs to the set $$\{E,Id\}\{L_a,R_b\}^*\{L_a, R_a\}^* 
\cup \{E,Id\}\{L_b,R_a\}^*\{L_b, R_b\}^*.$$
\end{proposition}

The proof, given at the end of the section, is a consequence of the next lemmas.

\begin{lemma}
\label{exL2.9}
Let $f$ be a morphism in $\{L_a,L_b,R_a, R_b \}^*$ different from the identity.
The morphism $f$ belongs to 
$\{L_a,R_b\}^*\{L_a, R_a\}^* \cup \{L_b,R_a\}^*\{L_b, R_b\}^*$ if
 and only if $f$ cannot be written $f = f_1f_2f_3$ with $f_1, f_3 \in
 \{L_a,L_b,R_a, R_b \}^*$ and $f_2$ verifying one of the four following properties:
\begin{enumerate}
\item $f_2 \in L_a \{L_a,L_b,R_a, R_b \}^\ast L_b \; \cup \;
L_b \{L_a,L_b,R_a, R_b \}^\ast L_a$, or
\item $f_2 = R_a g L_a$ with $g \not\in \{R_a,L_a\}^\ast$ or $f_2 = R_b g L_b$ with $g \not\in \{R_b,L_b\}^\ast$, or 
\item $f_2 \in R_a R_b^+ R_a$ or $f_2 \in R_b R_a^+ R_b$, or
\item $f_2 \in R_a^+ L_a^+ R_b = L_a^+ R_a^+ R_b$ or $f_2 \in R_b^+ L_b^+
R_a = L_b^+ R_b^+ R_a$.
\end{enumerate}
\end{lemma}

\begin{proof}
First we let the reader verify using Theorem~\ref{Lot02} that if $f$ belongs to 
$\{L_a,R_b\}^*\{L_a, R_a\}^* \cup \{L_b,R_a\}^*\{L_b, R_b\}^*$ then 
it cannot be written $f = f_1f_2f_3$ with $f_1, f_2, f_3$ as in the lemma.

From now on assume that $f$ cannot be written $f = f_1f_2f_3$ with
$f_1, f_2, f_3$ as in the lemma. Let $g_1, \ldots, g_n$ ($n \geq 1$
since $f$ is not the identity) in $\{L_a, L_b, R_a, R_b\}$ such that
$f = g_1\ldots g_n$.

We first consider the case where $g_1 = L_a$.  By Impossibility 1 for
$f_2$, for each $i > 1$, $g_i \neq L_b$. If there exists an integer $i
> 1$ such that $g_i = R_a$, then $g_1\ldots g_i = hL_aR_a^l$ or
$g_1\ldots g_i = hR_bR_a^l$ for a morphism $h$ and an integer $l \geq
1$.  In the first case by Impossibility 4 for $f_2$, for all integer
$j >i$, $f_j \neq R_b$. In the second case by Impossibilities 3 and 4
for $f_2$, for all integer $j >i$, we also have $f_j \neq R_b$. Thus
$f \in L_a\{R_b,L_a\}^\ast\{L_a,R_a\}^\ast$.

Assume now the more general case (than $g_1=L_a$) where there exists
an integer $i \geq 1$ such that $g_i=L_a$ and $g_j \neq L_a$ for $1
\leq j <i$ (the first occurrence of $L_a$ appears at the position
$i$). Samely as above, we show that $g=g_i \ldots g_n \in
L_a\{R_b,L_a\}^\ast\{L_a,R_a\}^\ast$.  By Impossibility 1 for $f_2$,
for each integer $j$, $1 \leq j < i$, $g_j \neq L_b$. Thus $g_j \in
\{R_a,R_b\}$ for each $1 \leq j <i$. We have three cases: If $f \in
R_a^\ast g$, then by Impossibility 4 for $f_2$, we have $f \in
L_a\{R_b,L_a\}^\ast\{L_a,R_a\}^\ast \cup \{R_a,L_a\}^\ast$.  If $f \in
h R_b^+R_a^\ast g$ for a morphism $h \in \{R_a,R_b\}^\ast$, then by
Impossibility 2 for $f_2$, $h \in R_b^\ast$ and so $f \in R_b^+
R_a^\ast g$; then by Impossibilities 3 and 4 for $f_2$ we have $f \in
\{L_a,R_b\}^\ast\{L_a,R_a\}^\ast$.  If $f \in R_b^\ast g$, $f \in
\{L_a,R_b\}^\ast\{L_a,R_a\}^\ast$.  So when there exists an integer $i
\geq 1$ such that $g_i=L_a$, $f \in \{L_a,R_b\}^\ast\{L_a,R_a\}^\ast$.

The case where there exists an integer $i \geq 1$ such that $g_i=L_b$ leads similarly to $f \in \{L_b,R_a\}^\ast\{L_b,R_b\}^\ast$.
 
Now we have to consider the case where for all $i$, $1 \leq i \leq n$, $g_i \not \in \{L_a,L_b\}$. Then by Impossibility 3 for $f_2$, necessarily, $f \in R_a^\ast R_b^\ast \cup R_b^\ast R_a^\ast$.
\end{proof}

\begin{lemma}\label{Form1}
Every morphism $f$ in $L_a \{L_a,L_b,R_a, R_b \}^\ast L_b \; \cup \;
L_b \{L_a,L_b,R_a, R_b \}^\ast L_a$ is strongly quasiperiodic.
\end{lemma}

\begin{proof}
We only prove the result for $f$ in $L_a \{L_a,L_b,R_a, R_b \}^\ast
L_b$ (the other case is similar exchanging the roles of the letters
$a$ and $b$).  Let $f=L_a f_1 f_2 \ldots f_n L_b$ with $n \geq 0$ and
$f_i$ in $\{L_a,L_b,R_a, R_b\}$ for all $1 \leq i \leq n$. We prove by
induction on $n$ that there exist morphisms $g$ and $h$ such that $f=g
L_a L_b h$ (and so from Lemma \ref{comp} and Fact~\ref{fact3}, $f$ is
strongly quasiperiodic).  The property is immediate for $n =
0$. Assume now $n \geq 1$.  If there exists $i$ between 1 and $n$ such
that $f_i=L_a$ or $f_i=L_b$, we can apply the induction hypothesis and Lemma~\ref{comp} to
conclude.  Now suppose that for all $i$, $f_i \not\in
\{L_a,L_b\}$. Three cases are possible:
\begin{itemize}
\item
if $f_1=R_a$, since $L_aR_a=R_a L_a$ from Theorem~\ref{Lot02}, $f =
R_a L_a f_1 \ldots f_n L_b$ and we conclude by the induction
hypothesis.
\item
If $f_n=R_b$ we can proceed similarly.
\item
Assume now $f_1=R_b$ and $f_n=R_a$ (this implies $n \geq 2)$.  Let $j$
be the greatest integer ($1 \leq j \leq n$) such that $f_j=R_b$. Then
$f=L_a f_1 \ldots f_{j-1} R_b R_a ^{n-j}L_b$, and by Theorem~\ref{Lot02}
$f=L_a f_1 \ldots f_{j-1} L_b L_a ^{n-j}R_b$. We
conclude by the induction hypothesis.
\end{itemize}
\vspace{-0.4cm}
\end{proof}

\medskip
Remark: we could have used another approach  observing that $L_a
R_b(w)$ ($L_aR_b(a) = aab$, $L_aR_b(b) = ab$) is $aba$-quasiperiodic
for every infinite word $w$ starting with $b$, and deducing that every
morphism of the form $L_a R_b f L_b$ with $f \in \{L_a,
R_a,R_b\}^\ast$ is strongly quasiperiodic.

\begin{lemma}\label{Form2}
Every morphism $f=R_a g L_a$ with $g \not\in \{R_a,L_a\}^\ast$ or
$f=R_b g L_b$ with $g \not\in \{R_b,L_b\}^\ast$ is strongly
quasiperiodic.
\end{lemma}

\begin{proof}
We only prove the first case, the other one is similar. Let $g=g_1
\ldots g_n$ (necessarily $n \geq 1$) such that $g \not\in \{R_a,L_a\}^\ast$
and for each $i$ between 1 and
$n$, $g_i \in \{L_a,L_b,R_a,R_b\}$.  If
there exists an integer $i$ such that $g_i=L_b$ then the result is
immediate from Lemma~\ref{Form1}.  Consequently we consider that $g
\in (\{L_a,R_a\}^\ast R_b)^+ \{L_a,R_a\}^\ast$. Thus the morphism $f$
can be decomposed $f=f_1 h f_2$ with $h \in R_a L_a^\ast R_b^+
R_a^\ast L_a$. If $i,j \geq 0, k \geq 1$ are the integers such that
$h=L_a^i R_a R_b^k L_a R_a^j$, Theorem~\ref{Lot02} shows that $h=L_a^i
L_a L_b^k R_aR_a^j$. Consequently Lemmas~\ref{comp} and 
\ref{Form1} imply that $h$ is strongly quasiperiodic.
\end{proof}

\medskip
Remark: here again we could have used another approach observing that $R_a
R_b(w)$ ($R_aR_b(a) = aba$, $R_aR_b(b) = ba$) is $aba$-quasiperiodic
for every infinite word $w$ starting with $a$, and deducing that every
morphism of the form $R_a R_b f L_a$ with $f \in \{L_a,
R_a,R_b\}^\ast$ is strongly quasiperiodic.

\medskip

This approach is used to prove:

\begin{lemma} 
\label{cor}
Any morphism $f$ in $R_a R_b^+ R_a \cup R_b R_a^+ R_b$
is strongly quasiperiodic.
\end{lemma}

\begin{proof}
Let $j \geq 1$ be an integer
such that $f = R_aR_b^jR_a$. Let $w$ be a word.  If $w$ starts with $b$, 
$R_b R_a(w)$ is $bab$-quasiperiodic, and so $f(w)$ is quasiperiodic.
If $w$ starts with $a$, $R_b^{j-1}R_a(w)$ also starts with $a$. Then
$R_aR_b^jR_a(a)$ is $aba$-quasiperiodic.
\end{proof}

\begin{lemma}\label{Form3}
Every morphism $f$ in $R_a^+ L_a^+ R_b =L_a^+ R_a^+ R_b$ or 
in $R_b^+ L_b^+ R_a = L_b^+ R_b^+ R_a$ is strongly quasiperiodic.
\end{lemma}

\begin{proof}
Theorem~\ref{Lot02} implies $R_a^+ L_a^+ R_b =L_a^+ R_a^+ R_b$ and
$R_b^+ L_b^+ R_a = L_b^+ R_b^+ R_a$.

We prove only the first case, the other one is similar.  Let $n \geq
1$.  It is easy to see that $R_a L_a^n R_b(w)$ ($R_a L_a^n R_b(a) =
aa^nba$, $R_a L_a^n R_b(b) = a^nba$) is $a^{n+1}ba$-quasiperiodic if
$w$ starts with $a$, and is $a^nbaa$-quasiperiodic if $w$ starts with
$b$. By Lemma~\ref{comp}, any morphism in $R_a^+ L_a^+ R_b$ is quasiperiodic.
\end{proof}

\medskip

\begin{proof2}{Proposition~\ref{caracStronglyQuasiperiodicMorphism}}

From Theorem~\ref{Lot02}, $EL_a=L_bE$ and $ER_a=R_bE$, so any Sturmian morphism can be written $fg$ with $f \in \{Id,E\}$ and $g \in  \{L_a,L_b,R_a,R_b\}^*$. Thus Proposition~\ref{caracStronglyQuasiperiodicMorphism} is a consequence of the following one: a morphism $f$ in $\{L_a,L_b,R_a,R_b\}^*$ is weakly quasiperiodic if and only if $f$ belongs to the set $X=\{L_a,R_b\}^*\{L_a, R_a\}^* \cup \{L_b,R_a\}^*\{L_b, R_b\}^*$. 

To prove this, assume first that $f \in \{L_a,L_b,R_a,R_b\}^*$ is weakly quasiperiodic.
By Lemma~\ref{comp}, this morphism cannot be written $f= f_1f_2f_3$ with $f_2$ a strongly quasiperiodic morphism. 
Hence by Lemmas~\ref{exL2.9}, \ref{Form1}, \ref{Form2}, \ref{cor} and \ref{Form3}, $f$ belongs to $X$.

Assume now that $f \in X$.
Since $f$ is Sturmian, it is quasiperiodic and so we just have to prove the existence of one word such that $f(w)$ is not quasiperiodic. So we just have to prove the existence of one word $w$ such that $f(w)$ is not quasiperiodic. We do it for $f \in \{L_a,R_b\}^*\{L_a, R_a\}^*$ (the
other case is similar). There exist morphisms $g \in \{L_a,R_b\}^*$
and $h \in \{L_a, R_a\}^*$ such that $f=gh$. We can verify that $h(aba^\omega)=a^nba^\omega$ for an integer $n \geq 1$, and so is a balanced word. By Lemmas~\ref{l4bis} and \ref{l4ter}, we thus deduce that $(g(h(aba^\omega))=f(aba^\omega)$ is not quasiperiodic.
\end{proof2}

\subsection{Weakly Sturmian morphisms on Sturmian words}

Proposition~\ref{caracStronglyQuasiperiodicMorphism} and Lemma~\ref{Form4} show that some
morphisms, as for instance $L_aR_a$, are weakly quasiperiodic whereas
they are strongly quasiperiodic on Sturmian words.  This section
allows us to characterize all these morphisms. Let us recall that
since a Sturmian morphism is quasiperiodic, any Sturmian morphism
is weakly or strongly quasiperiodic on Sturmian words.

\begin{proposition}
A Sturmian morphism different from $E$ and $Id$ is weakly
quasiperiodic on Sturmian words if and only if it belongs to
$\{E,Id\}\{L_a,R_b\}^* \cup \{E,Id\}\{L_b,R_a\}^*$.
\end{proposition}

\begin{proof}

Let us make a preliminary remark: for any morphism $f$, $f$ is weakly quasiperiodic on Sturmian words if and only if $Ef$ is weakly quasiperiodic on Sturmian words (since for any word $w$, $w$ is quasiperiodic if and only if $E(w)$ is quasiperiodic).

Assume first $f \in \{E,Id\}\{L_a,R_b\}^* \cup \{E,Id\}\{L_b,R_a\}^*$.
Without loss of generality, we can assume $f \in \{L_a,R_b\}^* \cup \{L_b,R_a\}^*$.
If $f$ belongs to $\{L_a,R_b\}^*$ (resp. to $\{L_b,R_a\}^*$), using Theorem~\ref{caracQuasiSturm} we observe that $f((L_aR_b)^\omega)$ (resp. $f((L_bR_a)^\omega)$) is not quasiperiodic. Since any Sturmian morphism is quasiperiodic, $f$ is weakly quasiperiodic on Sturmian words.

Now assume $f$ is weakly quasiperiodic on Sturmian words. 
Observe that from Theorem~\ref{Lot02}(2), $f \in \{E,Id\}\{L_a,L_b,R_a,R_b\}^*$. Without loss of generality, from the preliminary remark, we can assume that $f$ belongs to $\{L_a,L_b,R_a,R_b\}^*$ and prove that $f \in \{L_a,R_b\}^* \cup \{L_b,R_a\}^*$.
By Proposition~\ref{caracStronglyQuasiperiodicMorphism}, $f$ belongs to
$\{L_a,R_b\}^*\{L_a, R_a\}^* \cup \{L_b,R_a\}^*\{L_b, R_b\}^*$.
Assume by contradiction that $f \not \in \{L_a,R_b\}^* \cup \{L_b,R_a\}^*$.
One of the following four cases holds:
\begin{enumerate}
\item
$f = g L_aR_a$ with $g \in \{L_a,R_b\}^*\{L_a, R_a\}^*$;
\item
$f = g R_bR_a^i$ with $g \in \{L_a,R_b\}^*$, $i \geq 1$;
\item
$f = g L_bR_b$ with $g \in \{L_b,R_a\}^*\{L_b, R_b\}^*$;
\item
$f = g R_aR_b^i$ with $g \in \{L_b,R_a\}^*$, $i \geq 1$.
\end{enumerate}

Case 1: Assume $f= g L_aR_a$ and let $w$ be a non-quasiperiodic Sturmian word.  By
Lemma~\ref{Form4}, $f(w)$ is quasiperiodic. 

Case 2: Assume $f = g R_bR_a^i$ and let $w$ be a non-quasiperiodic
Sturmian word. By Theorem~\ref{caracQuasiSturm}, $w$ can be decomposed
over $\{L_a,R_b\}$ or over $\{L_b,R_a\}$.  So $f(w) =
gR_bR_a^iL_a(w')$ or $f(w) = gR_bR_a^iR_b(w')$ or $f(w) =
gR_bR_a^{i+j}L_b(w')$ for a (non-quasiperiodic) Sturmian word $w'$ and
an integer $j \geq 0$.  Thus by Lemma~\ref{Form4}, Lemma~\ref{cor} and
Lemma~\ref{Form2}, $f(w)$ is quasiperiodic.

Cases 3 and 4 are respectively similar to cases 1 and 2. In all cases, $f(w)$ is quasiperiodic for any non-quasiperiodic Sturmian word $w$, and so for any Sturmian word (by Fact~\ref{f(w)QP}). Thus $f$ is strongly quasiperiodic on Sturmian words. This is a contradiction, so $f \in \{L_a,R_b\}^* \cup \{L_b,R_a\}^*$.
\end{proof}

\section*{Acknowledgements}
Facts \ref{Lynd1} and \ref{overlap-free} were respectively observed by J.P. Duval and P. S\'e\'ebold during a talk given by the second author at the "Premi\`eres Journ\'ees Marseille-Rouen en combinatoire des mots" taking place in Rouen in June 2005. Many thanks to them and to J. Cassaigne, C. Mauduit and J.N\'eraud, the organizers of these "Journ\'ees" .

\bibliographystyle{plain}

\end{document}